\documentclass[a4paper,twocolumn,fontsize=10pt,DIV=16,abstract=true]{scrartcl}
\pdfoutput=1

\usepackage[utf8]{inputenc}
\usepackage[T1]{fontenc}
\usepackage[protrusion=true,expansion=true]{microtype}

\usepackage[backend=biber,alldates=edtf,seconds=true]{biblatex}
\usepackage{balance}

\usepackage{tikz}
\usetikzlibrary{arrows.meta,shapes}

\usepackage{caption}
\usepackage[skip=0pt]{subcaption}

\usepackage{todonotes}
\presetkeys{todonotes}{inline, caption={}}{}

\RequirePackage{booktabs}
\RequirePackage{longtable}
\RequirePackage{array}
\RequirePackage{tabularx}

\usepackage{amsmath}

\RequirePackage{siunitx}
\usepackage{scrlayer-scrpage}

\begin{document}
\title{A Glimpse of the Matrix (Extended Version)}
\subtitle{Scalability Issues of a New Message-Oriented Data Synchronization Middleware}
\date{}

\author{
Florian Jacob\\
\small Karlsruhe Institute of Technology\\
\small Institute of Telematics\\
\small florian.jacob@kit.edu
\and
Jan Grashöfer\\
\small Karlsruhe Institute of Technology\\
\small Institute of Telematics\\
\small jan.grashoefer@kit.edu
\and
Hannes Hartenstein\\
\small Karlsruhe Institute of Technology\\
\small Institute of Telematics\\
\small hannes.hartenstein@kit.edu
}

\maketitle

\begin{abstract}
  Matrix is a new message-oriented data synchronization middleware,
  used as a federated platform for near real-time decentralized applications.
  It features a novel approach for inter-server communication
  based on synchronizing message history by using a replicated data structure.
  We measured the structure of public parts in the Matrix federation
  as a basis to analyze the middleware's scalability.
  We confirm that users are currently cumulated on a single large server,
  but find more small servers than expected.
  We then analyze network load distribution in the measured structure
  and identify scalability issues of Matrix' group communication mechanism
  in structurally diverse federations.
\end{abstract}

\section{Introduction}
Matrix\footnote{\url{https://matrix.org}, \url{https://matrix.org/docs/spec/}} is a federated middleware for decentralized applications,
e.g.\ federated instant messengers\footnote{The name “Matrix” originates from the aim to reunite users on different communication channels by using Matrix as a meta-network to interconnect,
or \emph{“matrix together”}, previously segregated instant messaging platforms \cite{matrix-faq}.}.
It is based on a client-server architecture
where independent servers with limited mutual trust cooperate.
It provides topic-based publish-subscribe access on an eventually-consistent database
of messages and state changes.
Topics are called \emph{rooms} in Matrix.
A user's devices only connect to the user's Matrix \emph{homeserver}, which acts as representative for the user in the Matrix federation.
For each room, the participating servers form a communication group: They send messages to a room by broadcasting,
and receive messages and history from the other servers.
In contrast to other message-oriented middlewares, Matrix does not provide message passing but \emph{message history synchronization} \cite{matrix-spec}.
For applications, this has the advantage of not losing messages in transit, and of a consistent history between all of a user's devices \cite{matrix-spec}.
Messages and state changes form a partial order in the replicated per-room data structure
from which the current state is derived.

At present, the public federation is centered around one large server with about \num{50 000} daily active users as of January 2019.
However, Matrix is growing fast and is intended to be more decentralized.
Consequently, the structure of the network might change drastically in the future when more servers join the federation and users are distributed more evenly across them.
This will challenge the Matrix protocol in terms of scalability.
In addition to the public federation, at least one large, independent private federation exists:
In April 2019, the French government announced the beta release of its own, self-sovereign communication system based on
Matrix for the six million employees in the French public sector
in the form of a private federation between ministries~\cite{matrix-fosdem-french-state, lancement-de-tchap}.

In light of the increasing adoption of Matrix,
this paper focuses on examining the public Matrix federation as well as the protocol's scalability.
We crawled parts of the public federation and observed an imbalanced network.
We confirm that, despite its goal of decentralization, the public federation is mostly centralized in a single server, but also show that there are more small servers than expected.
Based on our measurements, we show scalability issues of the current message distribution algorithm.
We present ideas for a scalable replacement with the intention to allow leveraging Matrix' combination of messaging and storage in Internet-scale environments, e.g.\ in the Internet of Things.

The remainder of this paper is structured as follows.
First, we outline the architecture of Matrix (\autoref{sec:Architecture}) and describe its relation to other middlewares (\autoref{sec:RelatedWork}).
Then, we present our measurements of the public Matrix federation and discuss our results (\autoref{sec:Measuring}).
Finally, we elaborate on the scalability of the network (\autoref{sec:Scalability}),
and conclude by pointing out directions for future research (\autoref{sec:Conclusion}).

A peer-reviewed short version of this work is available at \url{https://doi.org/10.1145/3366627.3368106}.

\section{Architecture}
\label{sec:Architecture}
In this section, we provide a brief overview of the Matrix architecture based on the publicly available specification~\cite{matrix-spec}.
To use Matrix, an account on a \emph{homeserver} is required.
Homeservers act as a proxy for their users, as they execute actions and listen for reactions on their behalf.
The public federation is open, i.e. a user can either set up a personal homeserver
or register an account on a public homeserver.

In Matrix, all communication is organized in \emph{rooms}.
A room is not ``located'' on any single homeserver, but all homeservers that take part in a room are equal peers.
To become part of a room, a server contacts one of the servers known to be part of the room in question, allowing its users to join.

Messages sent to a room are expressed as \emph{events}.
Events can be categorized as either message events, like instant messages,
or state events, which update persistent information associated with a room, e.g. its name, access control policies and permissions.
The homeserver inserts new events into its copy of the \emph{Matrix Event Graph},
the distributed data structure that constitutes the core of Matrix,
and sends the event including additional meta data to all servers that participate in the corresponding room.
Each room has its own replicated Event Graph, independent of the graphs of other rooms.
Based on the exchanged information, the participating homeservers are able to reach an eventually consistent state.
The process of homeservers exchanging events for a room is called \emph{history synchronization}.
Eventual consistency \cite{eventual-consistency} represents a middle ground between weak consistency and strong consistency:
It guarantees that the distributed system will converge to a consistent state for a data object after a sufficiently long time without write accesses,
where the maximum time depends on factors like the number of participating servers, their load,
and the latencies between them.
In addition to the Event Graph itself forming a partial ordering on all events,
state events form a key-value store for the current persistent information associated with the room.
Participating servers independently execute a consensus algorithm on the partially ordered state events called the \emph{Matrix State Resolution Algorithm} to consistently agree on a room information state.

\section{Related Work}
\label{sec:RelatedWork}
\citeauthor*{ermoshina2016end} survey 30 instant messaging services including one based on Matrix \cite{ermoshina2016end}
that are either decentralized, end-to-end encrypted, or both.
Work on the Matrix middleware itself mainly targets
its end-to-end cryptography \cite{olm-crypto-review, johansen2017comparing},
while other aspects have yet to be examined.
The structure of other communication networks like the Internet Relay Chat (IRC) has been investigated before: \citeauthor*{deviants-in-irc} used an IRC crawler bot \cite{deviants-in-irc} similar to our crawler bot, but for gaining insights on the activities and the social graph of cyber criminals using IRC to stay in contact with their community.

Distributed variants of message-oriented middlewares~\cite{case_for_mom} like RabbitMQ~\cite{rabbitmq},
and distributed storage systems like Cassandra~\cite{cassandra},
are decentralized, i.e. every node operates indepentently.
Requests are equally distributed on all nodes, scaling linearly with the number of nodes.
This distribution requires that all nodes are controlled by a single entity,
i.e. it is not byzantine fault tolerant and can not be operated in a federation with limited trust.
XMPP\footnote{\url{https://xmpp.org/},\\ \url{https://xmpp.org/extensions/xep-0045.html}}
as message-oriented middleware supports public federation,
but its group communication mechanism relies on centralized, trusted
parties.
Matrix, however, is a promising new type of middleware for decentralized applications
as it combines distributed messaging and distributed storage, while supporting federations of limited trust.

Furthermore, we will show that all group communication mechanisms qualify as related work, be it broadcast, network-layer multicast, gossiping or other approaches.

\section[Measuring the Federation Structure]{Measuring the\\ Federation Structure}
\label{sec:Measuring}
\begin{figure}[htpb]
  \centering
  \begin{tikzpicture}[
    vertex/.style={draw, rounded corners, minimum width=1.5cm},
    user-room/.style={thick, solid},
    user-server/.style={thick, dashed},
    server-room/.style={thick, dotted},
  ]


  \node[vertex] (user) at (0,2.5) {User};

  \node[vertex] (room) at (6,2.5) {Room};

  \node[vertex] (server) at (3,1.9) {Server};

  \draw[user-server] (user) |- node[very near start, right,yshift=-1mm]{n} node[very near end, above,yshift=-.8mm]{1} (server);

  \draw[user-room] (user) to node[very near start, above]{n} node[very near end, above]{m} (room);

  \draw[server-room] (server) -| node[very near start, above,yshift=-.8mm]{n} node[very near end, left,yshift=-1mm]{m} (room);
\end{tikzpicture}
  \caption{Relations in the network structure}
  \label{fig:network_graph_relations}
\end{figure}
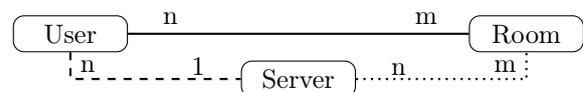

In this section, we describe our study on the structure of the public Matrix federation.
In order to model the relevant entities and their relationships, we define an abstract representation of the relationship between users, rooms and servers in form of a tripartite graph we call the \emph{network structure}, as shown in \autoref{fig:network_graph_relations}.
Each user node is connected to exactly one server node which is the user's homeserver (n:1),
a user node is connected to a room node if the user is a member of that room (n:m),
and lastly,
a server node is connected to a room node if at least one of its users is a member of that room (n:m).

In the following, we present our measuring method based on a crawler bot called \emph{DSN Traveller} (\autoref{sec:measuring:DSNTraveller}),
and provide details on our ethical considerations (\autoref{sec:measuring:Ethics}).
Finally, we present our findings and discuss possible conclusions based on our measurements (\autoref{sec:measuring:Results}).
The raw, anonymized network structure measurement is available for download \cite{matrix-data}.

\subsection{DSN Traveller Crawler Bot}
\label{sec:measuring:DSNTraveller}
The aim pursued with the DSN Traveller crawler bot
is to gather a partial snapshot of the network structure of the public Matrix federation by crawling public rooms.
To minimize interference with the network by the observation, a private Matrix homeserver was set up.
It was only used for hosting the bot and communications related to it.

For the crawling process, the publicly-available data from Matrix Voyager \cite{matrix-voyager-repo} was reduced to a room list and provided to the crawler bot.
After crawling, the graph was filtered with respect to special cases, ignoring the bot's homeserver and a subset of bridged nodes, which only act as stubs. 

\subsection{Ethics}
\label{sec:measuring:Ethics}
While the data we base our study on could be seen as publicly available and is accessible by everyone, no other entity is currently known that collects it systematically.
To balance the interest of Matrix users not getting tracked and the quantity of data that could be obtained,
users had the option to opt out on a per-room basis
by kicking or banning the bot's user from the room in question,
which happened 12 times over the course of the study.
On top of that, server operators were given the option of opting out with a whole server.
Two privacy-focused homeservers made use of this option.

The data gets pseudonymized directly after acquisition and filtering.
Deriving a deterministic identifier for each node is required during the crawling,
as users and servers that were already discovered in other rooms have to be deduplicated.
After collection, the pseudonyms get anonymized before the graph is stored on disk.

Furthermore, a number of measures were taken to inform the Matrix community about the crawler's activities and to ensure their benevolence and trust:
The bot had a website placed at the homeserver's domain, containing an explanation in layman's terms on what the bot was doing, why, and how one could interact with the bot \cite{dsn-traveller-website}.
A Matrix chat room \url{\#dsn-traveller:dsn-traveller.dsn.scc.kit.edu}
for questions or discussion about the bot's activities was operated.
The bot's appearance was also announced via This Week in Matrix (\url{\#twim:matrix.org}),
the official room for announcing and discussing news about Matrix-related projects, which are aggregated into a weekly blog post in the matrix.org blog \cite{twim}.
The announcement was done in This Week in Matrix 2018-05-25 and led to a number of discussions and questions about the project, but was perceived well in general \cite{twim2018-05-25}.
Finally, the code of the bot was published early on \cite{dsn-traveller-source}.

\subsection{Measurement Results}
\label{sec:measuring:Results}
We used the crawler bot to obtain a snapshot of the public Matrix federation's network structure in July 2018.
The bot observed a part of the public network: It joined \num{798} rooms, seeing \num{131 463} users on \num{2003} different servers.
The bot though knows about each user and homeserver that was present in at least one of the visited room.
Compared to Matrix Voyager's view at that time, the crawler bot
saw about two thirds of the servers.
As analyzing large tripartite graphs is challenging due to their inherent complexity,
the following analysis of the network structure focuses on each node group separately.

\emph{Server Group.}
Our measurements show that, from the \num{2003} homeservers in total,
by far the most servers have three or fewer users.
Only \num{15} servers were seen having more than \num{100} users.
The largest server had \num{76 271} users, the second had \num{37 751} users.
Together, this upper \SI{1}{\percent} of homeservers comprises \SI{87}{\percent} of the \num{131 463} total users found.
This means that most of the observed Matrix users are concentrated on very few homeservers,
which corresponds to the public perception of the system.
\begin{figure}[tbp]
\begin{subfigure}{\linewidth}
  \centerline{
    \resizebox{0.9\linewidth}{!}{
      \includegraphics{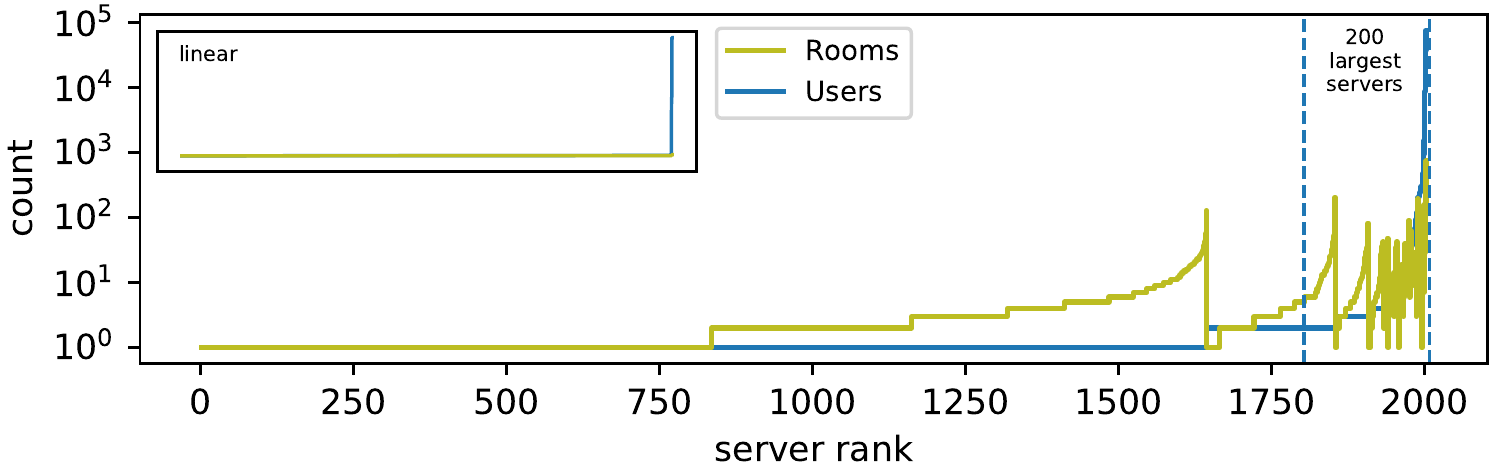}
    }
  }
  \caption{
    User and room count per server.
    Each particular server is assigned with a different, ascending rank
    based on its number of users.
  }
  \label{fig:per_server_count_step}
\end{subfigure}

\medskip{}
\begin{subfigure}{\linewidth}
  \centerline{
    \resizebox{0.9\linewidth}{!}{
      \includegraphics{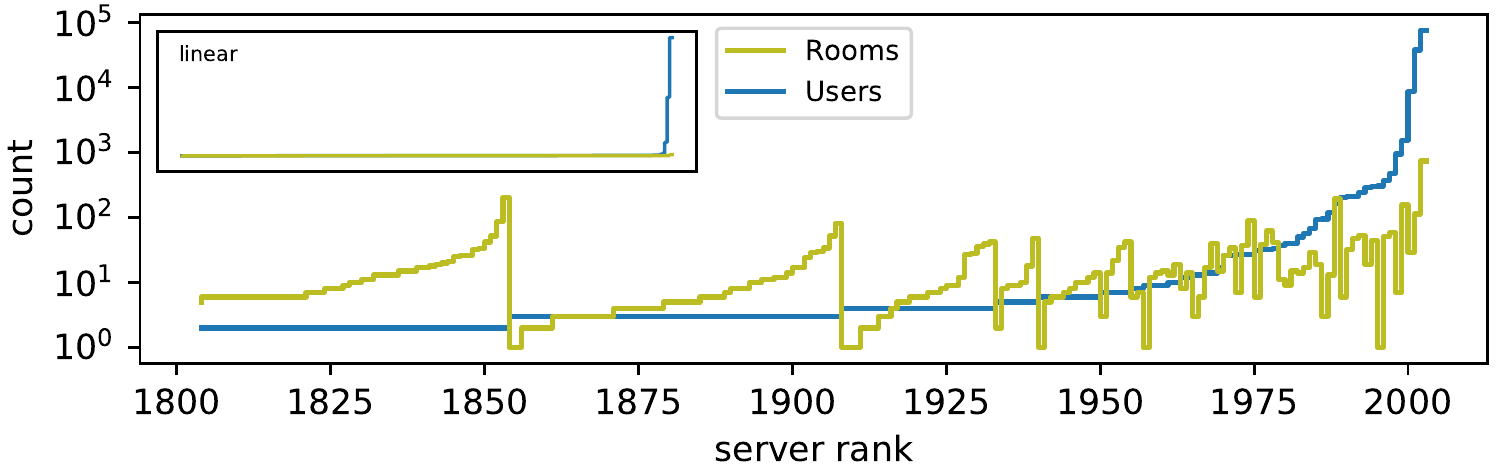}
    }
  }
  \caption{User and room count of the largest \num{200} servers only}
  \label{fig:per_server_count_step_zoomed}
\end{subfigure}

\caption[User and room count per server]{User \& room count per server on 2018-07-25}
\label{fig:per_server_count}
\end{figure}
\autoref{fig:per_server_count_step} shows the user and room count per individual server,
while \autoref{fig:per_server_count_step_zoomed} the largest \num{200} servers in more detail.
On the x-axis, servers are ranked by their number of users first and by their number of rooms second.
The y-axis depicts the number of users and rooms per server in log scale.
To remind the reader of the log scale, there are insets with the same plot in linear scale.
The plot exhibits a repeating spike pattern.
While the spike pattern itself is an effect of the second-order sorting,
the similar shape of all spikes is a noteworthy regularity:
For any number of users per server, there are servers with many and few rooms.

\begin{figure}[tbp]
\begin{subfigure}{\linewidth}
  \centerline{
    \resizebox{0.9\linewidth}{!}{
      \includegraphics{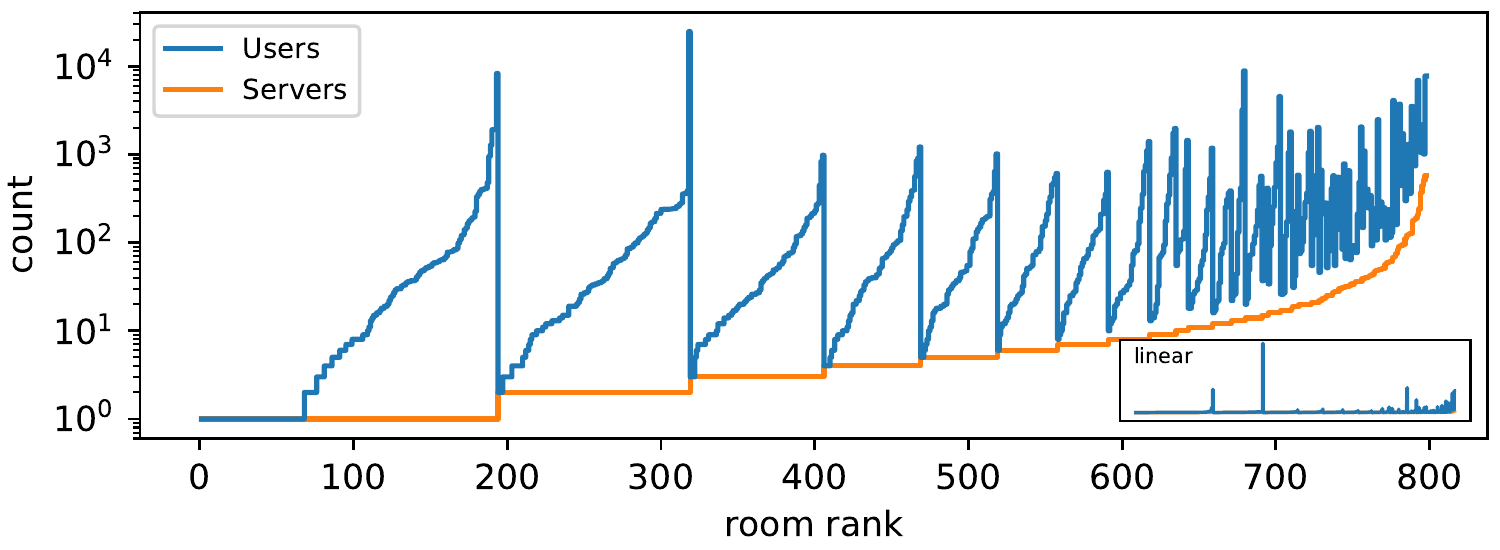}
    }
  }
  \caption{
    Server and user count per room.
    Each particular room is assigned with a different, ascending rank
    based on its number of servers.
  }
  \label{fig:per_room_count_step}
\end{subfigure}

\medskip{}
\begin{subfigure}{\linewidth}
  \centerline{
    \resizebox{0.9\linewidth}{!}{
      \includegraphics{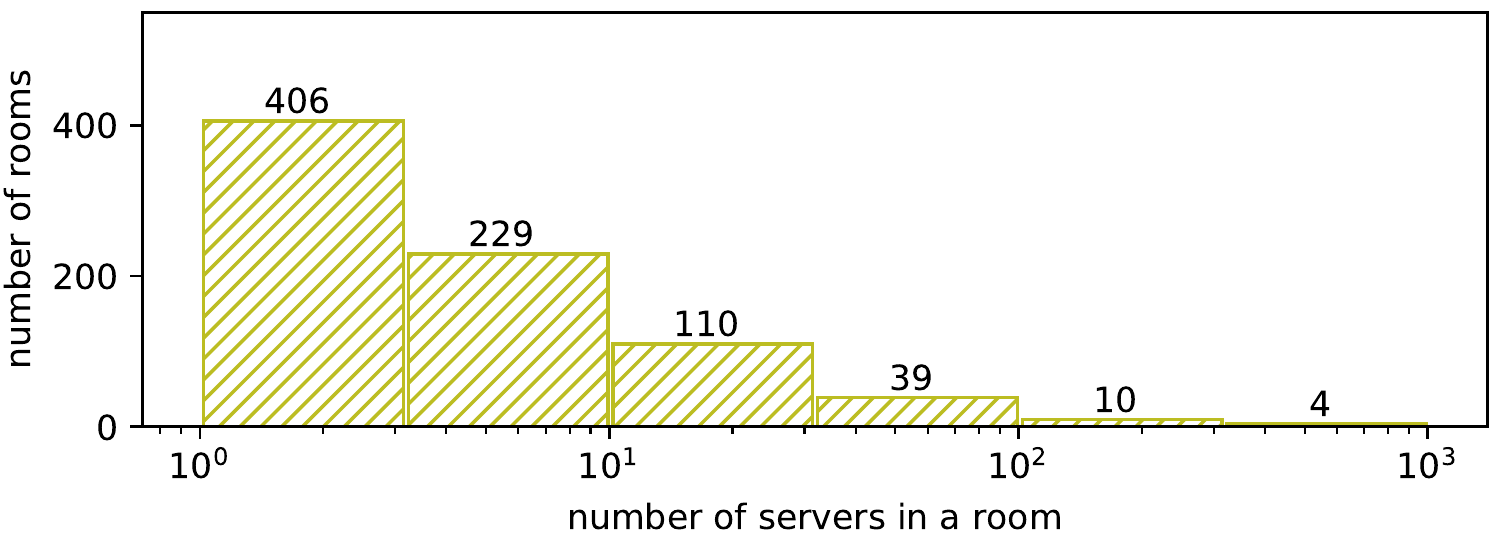}
    }
  }
  \caption{Histogram of servers per room; each bin is half an order of magnitude}
  \label{fig:per_room_count_server-hist}
\end{subfigure}

\medskip{}
\begin{subfigure}{\linewidth}
  \centerline{
    \resizebox{0.9\linewidth}{!}{
      \includegraphics{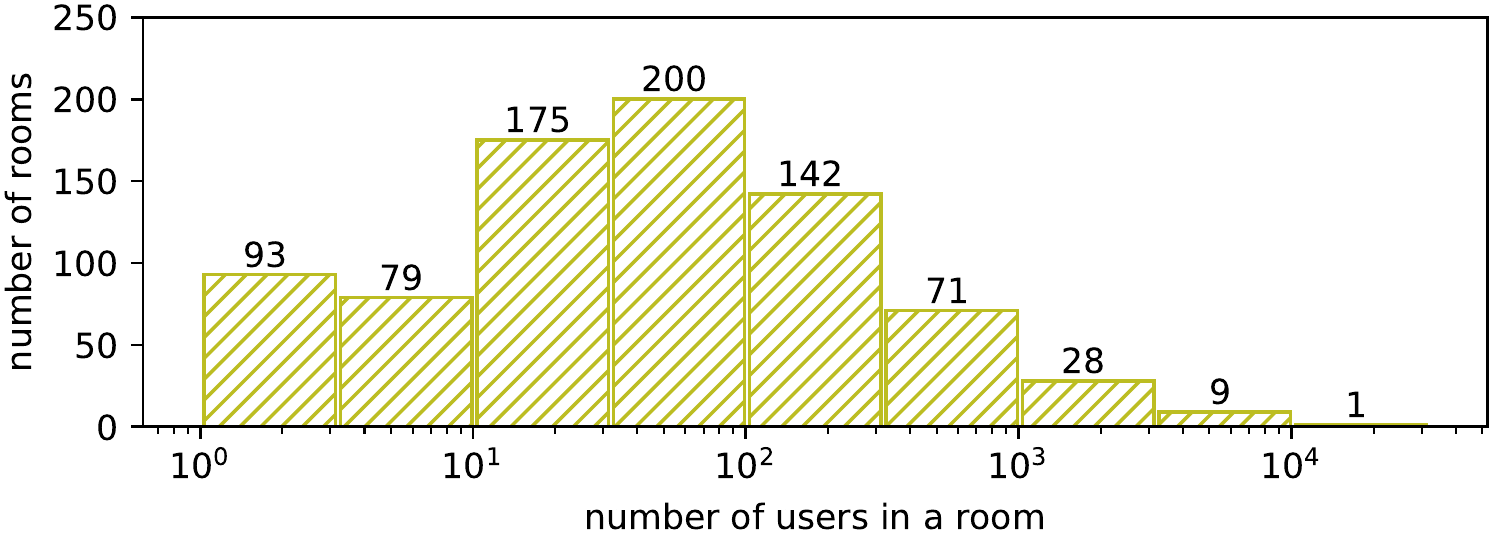}
    }
  }
  \caption{Histogram of users per room; each bin is half an order of magnitude}
  \label{fig:per_room_count_user-hist}
\end{subfigure}

\caption[Observed room composition]{Observed room composition on 2018-07-25}
\label{fig:per_room_count}
\end{figure}

\emph{Room Group.}
\autoref{fig:per_room_count} focuses on rooms:
The user and server count per individual room in \autoref{fig:per_room_count_step}
is ordered by number of servers first, and by number of users in second order.
It exhibits a spike pattern similar to the server-focused plots,
showing that a room with any number of servers can have only few or very many users.
As servers can only take part in rooms via users, a room cannot have more servers than users, and therefore there is a strong lower bound.
The server-wise largest room has \num{581} servers, which are \SI{76}{\percent} of all known servers.
Additionally, the number of servers and users per room are plotted as histograms,
using a logarithmic bin width of half an order of magnitude per bin.
The server histogram in \autoref{fig:per_room_count_server-hist}
shows that most rooms only have a few servers: \SI{83}{\percent} of all rooms have fewer than \num{10}.
The user histogram in \autoref{fig:per_room_count_user-hist} is of particular interest,
because its peak is in the range between \num{10} and \num{100} users, containing \SI{49}{\percent} of all rooms.
While the other histogram has its peak in the range between \num{1} and \num{10},
only \SI{22}{\percent} of all rooms have fewer than \num{10} users.
Still, while \SI{71}{\percent} of all rooms have \num{100} users or fewer,
there are rooms with many more users, up to the largest room with \num{24 729} users.
Rooms with the most servers do not have the most users, though:
Those in the rightmost bin have a maximum of \num{7756} users and a median of \num{1542} users.
This is caused by the user spikes in \autoref{fig:per_room_count_step} near room \num{200} and room \num{300} peaking an order of magnitude above rooms with the most servers.
Due to our measurement methodology, all private, invite-only rooms are missing in the data.
We conjecture that the peak between \num{10} and \num{100} in \autoref{fig:per_room_count_user-hist} is actually a result of this,
as we expect private rooms to be small (\num{1} to \num{10} users) and to outnumber the publicly visible rooms.

\emph{User Group.}
By far the most users were found in a single room only,
but a few users are in many rooms, up to the range of $10^2$.
\SI{94}{\percent} of the discovered users were only seen in three or fewer rooms.
The user who was seen in the most rooms was found in \num{207} rooms,
which corresponds to \SI{27}{\percent} of all rooms the crawler bot
visited.
We did not observe any crawler bot unknown to the Matrix community
that would have manifested as an omnipresent user.

All in all, our measurements confirm the assumed user concentration, while we found more small servers than expected.
The regularities in \autoref{fig:per_server_count} and \autoref{fig:per_room_count} allow the algorithmic generation of larger networks with similar characteristics to make predictions on Matrix' scalability.
The raw, anonymized measurement data is provided to the community \cite{matrix-data}.

\section{Scalability of the Network}
\label{sec:Scalability}
Matrix' group communication mechanism is inherently asymmetric between transmissions and receptions:
In a room with $n$ servers,
the sending server does $n-1$ individual transmissions to each receiving server.
We now explore the behaviour of the current group communication mechanism with the measured network structure.
We base our assessment on the network load, quantified by the number of incoming and outgoing event transactions between servers.

\subsection{A Model for Matrix}
\label{sec:scalability:model}
For analyzing the scalability of Matrix, we consider a reduced version of Matrix to explain the system's behaviour.
Our model operates on a frozen snapshot of a Matrix federation where all participants are honest and only send message events.
Therefore, servers do not need the State Resolution Algorithm.
The basic flow of events is modelled as a two-step process:
Users generate messages which target a specific room, and send them to their homeserver.
Servers are always reachable and serve an infinite number of events in parallel without processing delay.
To forward the event, the homeserver then sends a separate \emph{transaction} to all other homeservers which are part of the targeted room.
Transactions are put instantly on the wire, and servers do not need to wait for acknowledgements.
This implies that there is no batching of events into single transactions \cite{synapse-transaction-queue},
the model will therefore overestimate the number of transactions.
Modelling event batching requires timing information on transactions, which was not recorded for privacy reasons.

Additional assumptions are as follows:
The traffic pattern is determined by the average message rate per user $\lambda$, which we assume as fixed.
The room selection is modeled as uniform,
which means that a user sends messages to any room they are in with equal probability.

\subsection{Analytical Relations}
\label{sec:scalability:formulas}
Let $tx_{a \to b}$ be the number of transactions sent by $a$ to $b$
and $rx_{a \gets b}$ vice versa
for any two federating servers $a$ and $b$.
With $F$ being a set of federating servers, we define
$tx_s = \sum_{o \in F \setminus \{s\}} tx_{s \to o}$ and
and $rx_s = \sum_{o \in F \setminus \{s\}} rx_{s \gets o}$ to be the sum of all incoming and outgoing transaction for server $s$ in the federation $F$.
As each receiving server requires a separate transaction, we can state
$\sum_{s \in F} tx_s = \sum_{s \in F} rx_s$.

The number of transactions between servers depends on the network structure.
With $R_x$ being the set of all rooms that user or server $x$ is participating in,
$U_x$ being the set of all users that are part of server or room $x$,
we can express the average transactions rate of server $a$ to server $b$
as a function of the average per-user message rate $\lambda$ and the network structure:
\begin{align}
  tx_{a \to b} &= \sum_{r \in R_a \cap R_b} \sum_{u \in U_r \cap U_a} \frac{\lambda}{|R_u|} \label{formula:tx_ab}
\end{align}

In \autoref{formula:tx_ab}, one sums over all rooms that are on server $a$ as well as on server $b$;
all users in these rooms who are also users of server $a$ contribute with $\frac{\lambda}{|R_u|}$ events,
representing the assumption from \autoref{sec:scalability:model}
that every user sends messages with a rate of $\lambda$ equally distributed over all their rooms.
It follows that $rx_{a \gets b} = tx_{b \to a}$.

With $F_r \subseteq F$ being the federation subset of all servers participating in room $r$
and $Q_s = \{r \in R_s \mid |F_r| > 1\}$ being the set of all rooms
of server $s$ in which other servers participate, we can utilize \autoref{formula:tx_ab} to derive the network load of single servers with respect to their federation $F$:
\begin{align}
  \forall s \in F: tx_s &= \sum_{r \in Q_s}
    \sum_{u \in U_r \cap U_s} \frac{\lambda}{|R_u|} \cdot |F_r \setminus \{s\}| \label{formula:tx_s} \\
  \forall s \in F: rx_s &= \sum_{r \in Q_s}
    \sum_{u \in U_r \setminus U_s} \frac{\lambda}{|R_u|} \label{formula:rx_s}
\end{align}

The factor $|F_r \setminus \{s\}|$ in \autoref{formula:tx_s} is the key difference to \autoref{formula:rx_s},
which shows an asymmetry between sending and receiving.
This is caused by each server having to send a separate transaction to each receiving server,
multiplying the number of outgoing transactions:
The more foreign servers are in a room, and the more own users a server has in such a room, the more messages the server has to send.
But for the receiving side, the more foreign users are in the receiving server's rooms, the more messages it receives, regardless of how that users are distributed on foreign servers or how many foreign servers there are.
In the model, events are always sent individually (c.f. \autoref{sec:scalability:model}),
and incoming transactions are therefore independent of the number of foreign servers.
The number of transactions a server receives is only indirectly correlated with its own number of users,
as with more users on a server,
we can assume a higher probability of it being in more rooms with other servers.
The formulas were cross-checked with a Monte-Carlo simulation.

We can see that, due to the asymmetry,
moving from a centralized network to a more decentralized one will not improve the load distribution, but can actually make it worse: In the migration phase, if a user from a large server sets up his or her own homeserver, the large server will have to distribute the events of one fewer user, but instead has to distribute \emph{all} events from its \emph{remaining} users to the new homeserver, while that server only has to receive the relevant events for its user.
While the network load distribution will equalize when approaching a fully distributed state where every user has its own homeserver,
the network load for every server would then grow linearly in the number of users participating in a room,
as the scalability effect of reaching multiple users with single message to a server vanishes.

\subsection{Scalability Results}
In light of the possible network load distribution issues brought up using the formulas in \autoref{sec:scalability:formulas},
we now combine the formulas with the measurements of the network structure from \autoref{sec:measuring:Results} to show that the routing algorithm is already problematic for the measured network structure.

\begin{figure}[tbp]
\begin{subfigure}{\linewidth}
  \centerline{
    \resizebox{0.9\linewidth}{!}{
      \includegraphics{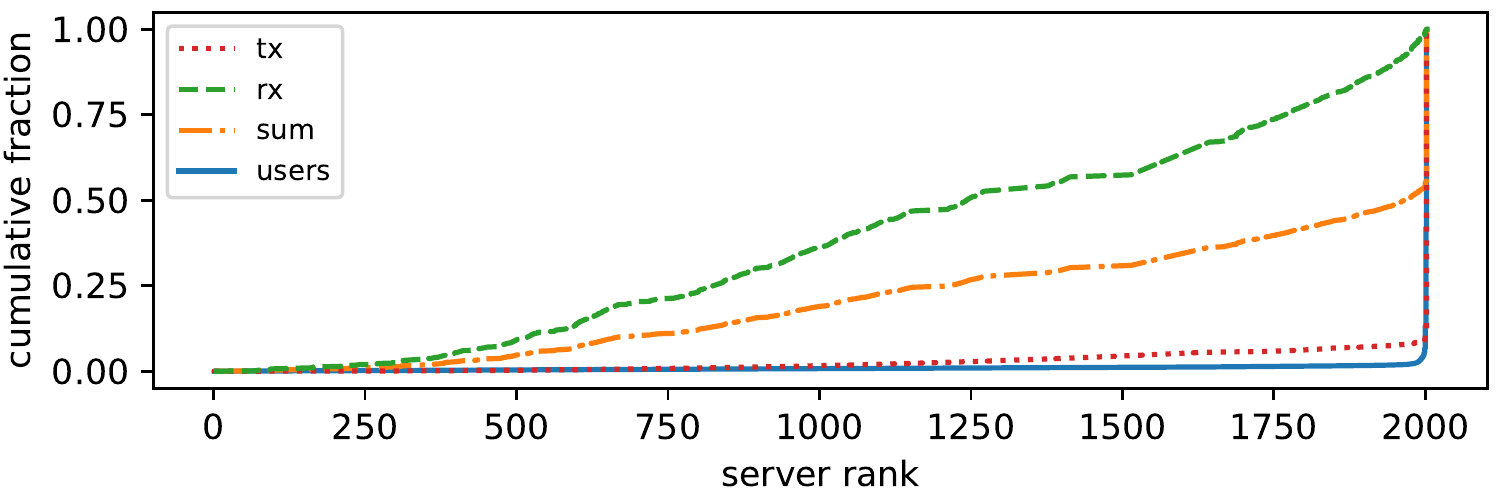}
    }
  }
  \caption{
    Cumulative fraction of network load per server.
    Each particular server is assigned with a different, ascending rank,
    based on its number of users.
    Due to all values being normed fractions, \texttt{sum} can be lower than \texttt{rx}.
  }
  \label{fig:tx_rx_relative_analytical_fractions}
\end{subfigure}

\medskip{}
\begin{subfigure}{\linewidth}
  \centerline{
    \resizebox{0.9\linewidth}{!}{
      \includegraphics{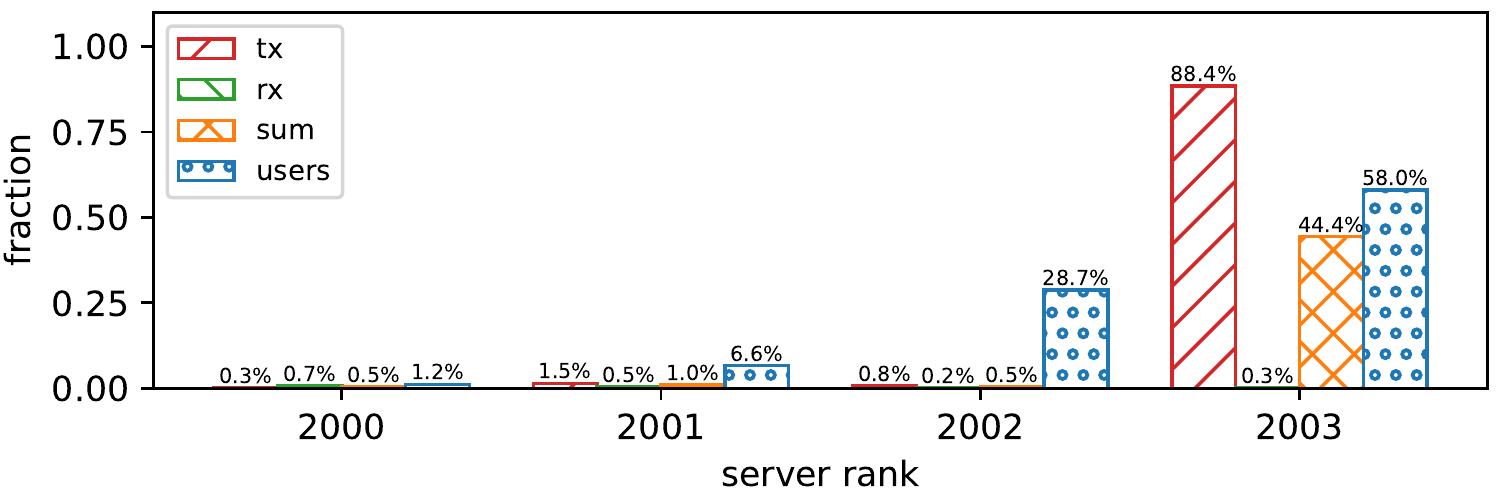}
    }
  }
  \caption{Network load of the largest four servers.}
  \label{fig:tx_rx_relative_analytical_largest-fractions}
\end{subfigure}

\caption{Cumulative fraction of network load per server}
\label{fig:tx_rx_relative_analytical}
\end{figure}

While it is evident from \autoref{formula:tx_s} and \ref{formula:rx_s} derived in \autoref{sec:scalability:formulas} that the number of users a server has correlates with the overall network transactions,
the actual relations are more complex and depend on the exact structure of the given network structure.
\autoref{formula:tx_s} and \ref{formula:rx_s} allow us to calculate the transmitted (\texttt{tx}) and received (\texttt{rx}) as well as the total number of transactions (\texttt{sum}) for any server,
as displayed in \autoref{fig:tx_rx_relative_analytical_fractions}.
To evaluate the results in a meaningful way, the form of cumulative fractions was used:
All measurement values are normalized to the population size,
and each value is added onto the sum of the previous measurements.
This is similar to an empirical distribution function.
As the number of received transactions equals the number of sent transactions (see \ref{sec:scalability:formulas}), the relation between \texttt{rx} and \texttt{tx} also translates to their absolute values.
To put the numbers into perspective, the cumulative fraction of users is plotted as well,
and the individual servers are ordered by their number of \texttt{users} first, and by \texttt{tx} in second order.
In the center of the plot, from 750 to 1500, the $tx$, $rx$ and $sum$ curves rise with a roughly constant incline.
This shows that those servers send and receive a similar amount of messages to and from the federation,
and are involved in a similar number of the total transactions.
The very steep rise near the 2000th server shows that,
while the servers left of it almost receive \SI{100}{\percent} of all received messages,
they only send about \SI{10}{\percent} of all sent messages.
In sum, those servers are part of just above \SI{50}{\percent} of all traffic.
The few remaining servers therefore make up what is missing to \SI{100}{\percent}:
They almost receive no transactions, but send about \SI{90}{\percent}.
In total, they are part of just below \SI{50}{\percent} of all traffic.

As the constituents of the steep rise
at the servers with the highest rank cannot be differentiated in \autoref{fig:tx_rx_relative_analytical_fractions},
\autoref{fig:tx_rx_relative_analytical_largest-fractions} only shows the numbers for the largest four servers in the form of a non-cumulative bar diagram.
This shows that the single, largest server is involved in \SI{44.5}{\percent} of all messages,
sends \SI{88.4}{\percent}, but receives only \SI{0.6}{\percent} of all messages.
The other three servers are only part of a much smaller fraction of the overall traffic.
Therefore, the steep rise is actually caused by a \emph{single} server.
Whereas receiving transactions is distributed across all servers,
sending transactions as well as the accumulated load of sending and receiving is centralized to a single server, which is involved in \SI{44.5}{\percent} of all messages.

The reason for the observed load centralization lies in unequally distributed users:
Most events are generated on large servers, which then have to be distributed to many small servers,
while small servers can reach many users with a single transmission to a large server.
Decreasing the degree of centralization will worsen the load distribution during the transition phase
when more users set up small servers but while the majority is still cumulated.
When reaching full decentralization (i.e. each user is represented by a separate homeserver),
the efficiency benefit of reaching multiple users with a single transmission vanishes.
These aspects affect the scalability of Matrix and hinder the future growth of the public federation.
At present, Matrix is mainly used for communication between humans, but the middleware explicitly targets Internet of Things communication \cite{matrix-faq}. Such a machine-to-machine use case potentially increases the number of involved parties and message frequencies by several orders of magnitude, and aggravates the problem of limited scalability in the group communication mechanism.
Due to the limited trust between servers,
load balancing can not be achieved by moving users away from busy servers as with non-federated middlewares.
For a decentralized Matrix network, a scalable group communication mechanism has to evenly distribute the Price of Anarchy, i.e. the loss of system efficiency compared to a central authority making all decisions \cite{price-of-anarchy}.
This poses a combined optimization problem of communication topology and routing mechanism.
As each Matrix room is an independent federation which can differ vastly in structure even if the same servers take part,
we envision to use a room structure adaptive mechanism:
For a given room, participating servers know which other servers participate and with how many users they take part.
Given this information, servers can agree on a suitable per-room group communication mechanism, to deliver Matrix' promising combination of federation, messaging and storage to decentralized applications at any scale.
This means that the same servers might use traditional broadcasting in a room with a few servers only, achieving optimal end-to-end latency,
but use gossipping in another room where many servers with few users take part, or even IPv6 multicast if all participating servers have a suitable interconnect.
This is why all forms of group communication are considered related work.

Every mechanism is optimized for different topologies, and via the unifying Event Graph concept,
Matrix has the capability of acting as an integration platform for any form and number of routing mechanisms run in parallel to each other,
without any changes to the layers on top of Matrix.

\section{Conclusion \& Future Work}
\label{sec:Conclusion}
We presented a measurement of the network structure of the public Matrix federation.
The crawler bot and the raw, anonymized data is provided to the community.
We identified scalability issues in the group communication mechanism of the Matrix middleware
in form of load centralization in structurally diverse federations,
which cannot be mitigated by rebalancing users due to the limited trust between Matrix servers.
For future research, we aim at investigating different group communication mechanisms in the context of Matrix, pursuing the goal of a room-structure-adaptive algorithm.
Such an algorithm would choose a suitable group communication mechanism for a given room structure to address the scalability issues on the horizon.
To test the algorithms' scalability,
we will model and simulate larger federations with similar characteristics using the regularities in the presented measurements.
We will continue to observe the evolution of the public Matrix federation towards more decentralization.

\balance
\printbibliography

\end{document}